\documentclass[12pt]{article}

\usepackage{newtxtext,newtxmath}

\usepackage{graphicx}

\usepackage[letterpaper,margin=1in]{geometry}

\renewenvironment{abstract}
	{\quotation}
	{\endquotation}

\date{}

\makeatletter
\renewcommand{\fnum@figure}{\textbf{Figure \thefigure}}
\renewcommand{\fnum@table}{\textbf{Table \thetable}}
\makeatother

\usepackage{scicite}

\usepackage{url}

\newcommand{\CSS}{Co$_3$Sn$_2$S$_2$}
\newcommand{\bm}[1]{\boldsymbol{#1}}

\def\scititle{
General \textit{ab initio} framework for electronic-order-induced lattice-dynamics symmetry breaking}

\title{\bfseries \boldmath \scititle}

\author{
	Shuai Zhang$^{1}$,
    Mengqi Wang$^{1}$,
    Pan Zhang$^{1}$,
    Tiantian Zhang$^{1\ast}$\and
	\small$^{1}$Institute of Theoretical Physics, Chinese Academy of Sciences, Beijing 100190, China.\and
	\small$^\ast$Corresponding author. Email: ttzhang@itp.ac.cn\and
}

\begin{document} 

\maketitle

\begin{abstract} \bfseries \boldmath
Conventional \textit{ab initio} approaches are unable to describe phonon time-reversal symmetry ($\mathcal{T}$) breaking.
Here, we develop an \textit{ab initio} framework, grounded in molecular Berry curvature (MBC) theory, that captures electronic-order-driven symmetry breaking in lattice dynamics. 
Using Co$_3$Sn$_2$S$_2$ as a model system, our \textit{ab initio} framework yields phonon spectra that break both $\mathcal{T}$ and mirror symmetries, quantitatively reproduce the observed phonon splittings observed in experiments, and reveal distinct microscopic origins for the $E_g$ and $E_u$ modes: $E_g$ splitting is governed by MBC and is accurately captured by our algorithm, whereas $E_u$ splitting is enhanced by the Fano resonance and matches the experimental data once the Fano-factor correction is included. Leveraging this algorithm, we predict several candidate materials with nonzero electronic-order-driven symmetry breaking in lattice dynamics, establishing a first-principles route to understand electron-phonon coupling, phonon magnetism, and related Hall-type lattice responses.
\end{abstract}

\noindent
\subsubsection*{Short title: First-principles reveal phonon magnetism and $\mathcal{T}$ breaking}

\subsubsection*{Teaser: First-principles calculations reveal effective phonon magnetism and time-reversal symmetry breaking induced by electronic order.}

\section{Introduction}

Within Landau's theory, electronic orders, such as {magnetic/}spin, charge, and orbital order, are typically characterized by local order parameters and strong electron interactions, and are invariably accompanied by spontaneous symmetry breaking. This contrasts with phonon systems, where such symmetry breaking is rare. 
Recent work has shown that time-reversal symmetry ($\mathcal{T}$) breaking occurs in phonons by coupling to external fields~\cite{Baydin2022_PRB, Cheng2020_Cd3As2, Luo2023_CeF3_Science, Chaudhary2024_PRB, Wu2023_Fe2Mo3O8_NP, David2024_CoTiO3_PNAS,  Nova2017_ErFeO3_NP, Schaack1977_CeCl3, CeCl3_2022,  Niu2021_phononMag, Xue2025_Extrinsic}, spins~\cite{Ren2024_PH_SPIN_PRX, Liu2021_Magnon_phonon} and chiral excitations~\cite{FeGeTe_2D_chiral_phonon_Du_2019}. {Moreover,} phonon energy splitting in ferromagnetic~(FM) Weyl semimetal \CSS{} directly demonstrates this effect~\cite{CoSnS_CP_2025, CoSnS_PRL_2025}. 
Although \textit{ab initio} methods can readily describe electronic symmetry breaking, such as $\mathcal{T}$ in magnetic systems, {this} concomitant symmetry breaking in the phonon system induced by {FM} electronic order remains inadequately explored.
Thus, a quantitative understanding of symmetry breaking in lattice dynamics from first principles~\cite{JunRen_Shi2017,Bistoni2021_Noncollinear_PRL,Bonini2023_CrI3_TRS, Ren2024_PH_SPIN_PRX,HJZhang2025_EPC, chen2025magicnonlocalgeometricforce} is essential, and it is crucial for interpreting phonon-related transport~\cite{Strohm2005_THE, Nagaosa2010_THE_QuantumMagnets, Ideue2017_THE_multiferroics, Zhang2021_ATHE_VI3, Nagaosa2019_Berry_THE,
Niu2022_MolecularBerry, Xue2025_Extrinsic} and optical properties~\cite{CoSnS_CP_2025,CoSnS_PRL_2025,nagaosa2024nonreciprocal,CeCl3_2022}.

In this work, we develop an \textit{ab initio} algorithm based on molecular/nuclear Berry curvature (MBC) theory~\cite{MT1979_lattice_dyn, Niu2022_MolecularBerry, NBC_PRB_2022} to capture the magnetism-induced symmetry breaking in lattice dynamics.
It necessitates the inclusion of spin-orbit coupling (SOC) when the magnetic order originates from spin~\cite{Ren2024_PH_SPIN_PRX}. In the absence of SOC, $\mathcal{T}$ can be broken only through the orbital (spatial) part of the wave function, as in loop-current order~\cite{Varma_copper_PRB_1997, Varma_copper_PRL_1997}. Consequently, circularly polarized phonon splitting provides an effective probe for detecting loop-current order~\cite{Grissonnanche2020_NP, Chen_loop_CP_PRL2025}.

Using Co$_3$Sn$_2$S$_2$ as a representative system, we demonstrate that the MBC contribution breaks both $\mathcal{T}$ and mirror symmetries in the lattice dynamics when both the ferromagnetic order and SOC are considered. We further show that the breaking of both symmetries is essential for a complete theoretical description of lattice dynamics, giving rise to phonon splittings that extend throughout the Brillouin zone (BZ). Our algorithm can be used to predict candidate materials exhibiting phonon splitting or effective phonon magnetism, several of which are identified in this work. We find that while phonon splitting or effective phonon magnetism requires phonon modes with nonzero angular momentum, the magnitude of the effect emerges further from the {interplay} of magnetic order, SOC, and electron-phonon interactions.

\section{Results}

\subsection{Lattice dynamics without MBC}

\begin{figure*}
    \centering
    \includegraphics[width=1.0\textwidth]{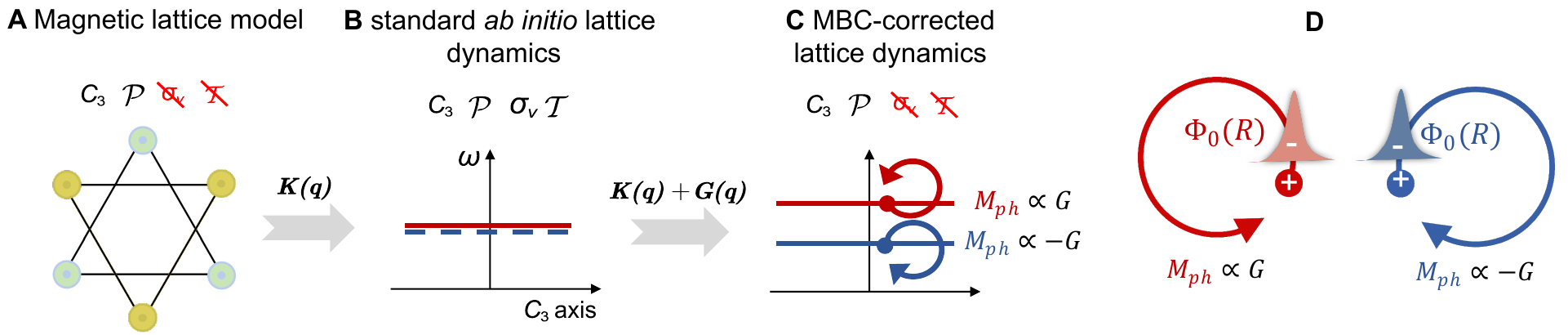}
    \caption{
\textbf{Schematic illustration of lattice dynamics in magnetic materials under the standard \textit{ab initio} approach and the MBC-corrected framework}.
(A) The model lattice preserves $C_3$, inversion ($\mathcal{P}$), vertical mirror ($\sigma_v$), and time-reversal ($\mathcal{T}$) symmetries. {Collinear} ferromagnetic order  aligned along the $C_3$ axis and parallel to the vertical mirror plane breaks both $\sigma_v$ and $\mathcal{T}$. 
(B) Phonon spectra obtained by standard \textit{ab initio} calculations, where the doubly degenerate phonon dispersion is enforced by $\sigma_v$ and $\mathcal{T}$ in the dynamical matrix $K(\bm{q})$.
(C) By introducing the $G(\bm{q})$ into the lattice dynamics, our algorithm accurately captures the phonon spectra of magnetic materials. The MBC term generates opposite effective phonon magnetic moments $M_{ph}$ for right- and left-handed rotational vibrations, thereby breaking $\sigma_v$ and $\mathcal{T}$ symmetries.
(D) Schematic illustration of the MBC arising from the adiabatic evolution of the electronic ground state $\Phi_{0}$ under different phonon modes. Phonon modes with opposite circular polarizations acquire MBC corrections of opposite sign.}
    \label{fig:Dyn_MBC}
\end{figure*}

In conventional first-principles calculations, the dynamical matrix $K_{ij}$ is constructed to preserve time-reversal symmetry $\mathcal{T}$, even when this symmetry is broken by the magnetic order, leading to inaccurate phonon spectra. For instance, the lattice model shown in Fig.~\ref{fig:Dyn_MBC}A possesses $C_3$ rotational, inversion ($\mathcal{P}$), vertical mirror ($\sigma_v$), and $\mathcal{T}$ symmetries. Collinear ferromagnetic order parallel to the rotational axis {and mirror plane} breaks both $\mathcal{T}$ and $\sigma_v$.
However, conventional first-principles approaches still enforce these symmetries in the dynamical matrix (see Supplementary Materials), yielding phonon spectra that preserve the same symmetries as the nonmagnetic phase, as shown in Fig.~\ref{fig:Dyn_MBC}B. The accurate phonon spectra, which reflect the true lattice dynamics of the magnetic system, should instead correspond to Fig.~\ref{fig:Dyn_MBC}C and lie beyond the reach of conventional first-principles methods.
In the following, we show how MBC corrections modify lattice dynamics and develop an \textit{ab initio} algorithm incorporating these effects. Applying this framework to Co$_3$Sn$_2$S$_2$ as a representative system, we demonstrate that it captures the microscopic origin of the experimentally observed phonon splittings. 

\begin{figure*}
    \centering
    \includegraphics[width=\textwidth]{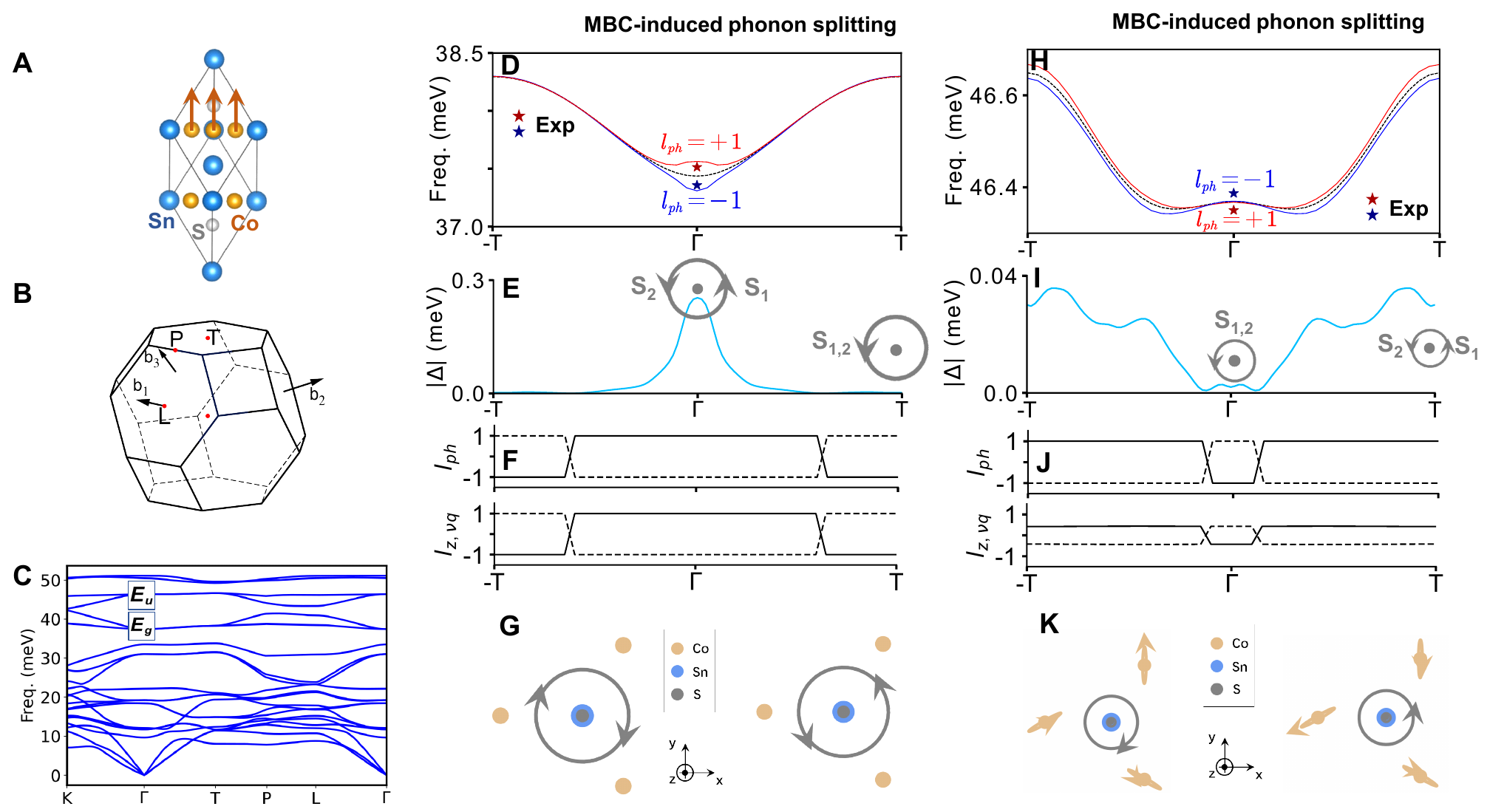}
    \caption{
    \textbf{MBC-induced phonon splitting for $E_g$ and $E_u$ modes in Co$_3$Sn$_2$S$_2$}.
    (A) Crystal structure of \CSS{}. Brown arrows indicate the local magnetic moments on Co atoms. The magnetic order breaks the $\sigma_v$ and $\mathcal{T}$ symmetries, but keeps the $C_3$ and $\mathcal{P}$ symmetries. (B) Brillouin zone of \CSS{}. (C) Phonon spectra of FM-ordered \CSS{} with SOC, excluding the molecular Berry curvature contribution. The coordinate of $K$ is ($\frac{1}{3}$, 0, -$\frac{1}{3}$){/} ($\frac{1}{3}$, $\frac{1}{3}$, 0) in the basis of the primitive/{conventional} lattice vectors. 
    (D) Phonon dispersion of the $E_g$ mode along the $C_3$-invariant path $-\text{T} \leftrightarrow \Gamma \leftrightarrow  \text{T}$, calculated without MBC (black dashed line) and with MBC (blue and red solid lines). The colored stars mark the energy positions of the $E_g$ mode observed experimentally.  
    (E) Energy splitting of the $E_g$ mode along the $C_3$-invariant path.
    The insert figure shows the vibrational modes of the S atoms at the $\Gamma$ and T points. 
    (F) Evolution of the pseudo-angular momentum ($l_{ph}$) and the $z$-component of the angular momentum ($l_{z,\nu \bm{q}}$) along the $C_3$-invariant path. {The solid (dashed) line corresponds to the higher-frequency (lower-frequency) phonon branch shown in (D), and their intersection indicates a band-crossing point.}
    (G) Atomic displacement patterns of the two $E_g$ modes at the $\Gamma$ point. (H)--(K) are the corresponding contents for the $E_u$ mode.}
    \label{fig:crystal_band_phonon}
\end{figure*}

\CSS{} is the first experimentally confirmed magnetic Weyl semimetal~\cite{Xu2018_PRB_CoSnS_Weyl, Liu2018, Wang2018_NC,Yin2019_CSS_Negative, science.aav2873,science.aav2334,YAN201857,PhysRevLett.124.077403,PhysRevMaterials.4.044203,Xu2020,Okamura2020_magopt_CSS} with a centrosymmetric space group of $R\bar{3}m$ ($\mathcal{P}$-preserving), thus the phonon modes can be classified in Raman active ones and {infrared} (IR) active ones by the $\mathcal{P}$ parity at $\Gamma$. 
Figure~\ref{fig:crystal_band_phonon}A shows the primitive cell of \CSS{} with brown arrows indicating the local magnetic moment.
Below $T_{\rm C} \sim 175$ K~\cite{Liu2018}, {collinear} FM order aligned along the $C_3$ axis and parallel to the vertical mirror plane breaks both $\mathcal{T}$ and vertical mirror symmetries~\cite{Mag_CSS_PRB}, enabling an SOC-essential Weyl semimetal phase, reducing the point group from $D_{3d}$ to $C_{3i}$ (magnetic space group $R\bar{3}m'$). 

Figure~\ref{fig:crystal_band_phonon}C shows the phonon spectra of FM Co$_3$Sn$_2$S$_2$ with SOC, obtained from standard \textit{ab initio} calculations, where the two experimentally observed $E_{g}$ and $E_u$ modes remain degenerate, analogous to Fig.~\ref{fig:Dyn_MBC}B. Thus, the dynamical matrix from the standard first-principles calculation still preserves $\mathcal{T}$ and $\sigma_v$~\cite{Zhang2025_CP}, despite the {collinear} FM order breaking them. 
However, the experimental data are non-degenerate, analogous to Fig.~\ref{fig:Dyn_MBC}C. 

\subsection{MBC modified lattice dynamics}

To capture lattice dynamics in magnets accurately, the adiabatic electronic evolution must be included~\cite{EM_THE_Qin2011_PRL,Niu2022_MolecularBerry}, as it feeds back into the lattice motion and yields the MBC, $G(\bm{q})$, as shown in Figs.~\ref{fig:Dyn_MBC}C and D.
The equation of motion~(EOM) reads:
\begin{equation}
\label{eq:EOM}
    [\Tilde{K}(\bm{q})+{2}\tilde{G^\dagger}(\bm{q})\Tilde{G}(\bm{q}) {+ 2i \omega_{\nu\bm{q}} \Tilde{{G}} (\bm{q}) } ] \epsilon_{\nu \bm{q}}= \omega^2_{\nu \bm{q}}\epsilon_{\bm{q}},
\end{equation}
where $\epsilon_{\nu\bm{q}}$ is the polarization vector, $\Tilde{K}_{\kappa \alpha, \kappa'\beta}(\bm{q})=\frac{1}{\sqrt{M_{\kappa}M_{\kappa'}}}K_{\kappa \alpha, \kappa'\beta}(\bm{q})$ and $\Tilde{G}_{\kappa \alpha, \kappa'\beta}(\bm{q})=\frac{1}{2\sqrt{M_{\kappa}M_{\kappa'}}}G_{\kappa \alpha, \kappa'\beta}(\bm{q})$ are the mass-weighted dynamic matrix and MBC. $\kappa$ is the index of atom in the primitive cell, and $M_{\kappa}$ is the corresponding atom mass, $\alpha$, $\beta$ $\in \{x, y, z\}$.
To solve this equation, we can take $\Tilde{G}(\bm{q})$ as the {first-order} perturbation, and discard the second-order perturbation $\tilde{G^\dagger}(\bm{q})\Tilde{G}(\bm{q})$. The zero-order equation reads:
\begin{equation}
      {\Tilde{K}(\bm{q})\epsilon^{(0)}_{\nu \bm{q}}= \omega^{(0)2}_{\nu \bm{q}}\epsilon^{(0)}_{\bm{q}}},
\end{equation}
and the first-order modification of $\omega^{2}_{\nu \bm{q}}$ is obtained by diagonalizing the matrix
$\langle \epsilon_{\nu\bm{q}} | 2 i \omega^{(0)}_{\nu\bm{q}} \Tilde{G}_{\kappa \alpha, \kappa'\beta}(\bm{q})|\epsilon_{\nu'\bm{q}} \rangle$, where $\nu=\nu'$ for the non-degenerated case. Detailed discussions are in the Supplementary Materials.

Experiments rule out resonant magnon-phonon coupling in \CSS~\cite{Liu2020spin}, allowing it to be safely excluded for an accurate description of the FM-ordered lattice dynamics~\cite{Ren2024_PH_SPIN_PRX, Bonini2023_CrI3_TRS}. 
Furthermore, Ref.~\cite{CoSnS_CP_2025} shows that Co$_3$Sn$_2$S$_2$ is a poor metal, with an optical conductivity on the order of $10^3~\Omega^{-1}\,\mathrm{cm}^{-1}$. Moreover, Ref.~\cite{CoSnS_PRL_2025} reports that Co$_3$Sn$_2$S$_2$ exhibits a clear Raman spectrum, in contrast to conventional metals, {thus} the Born--Oppenheimer approximation (BOA) remains valid in this material.
Although the MBC was proposed to break $\mathcal{T}$-symmetry~\cite{Niu2022_MolecularBerry,Niu2022_MolecularBerry,Ren2024_PH_SPIN_PRX}, it alone is insufficient to lift the degeneracy of phonon bands along $-\text{T} \leftrightarrow \Gamma \leftrightarrow \text{T}$ due to the presence of $\sigma_v$~\cite{Zhang2025_CP}. 
Therefore, the MBC-induced $\sigma_v$ breaking is essential for explaining the phonon splitting in experiments.

\subsubsection{MBC-modified $E_g$ mode}

Figure~\ref{fig:crystal_band_phonon}D shows the $E_g$-mode phonon spectra obtained from our MBC-based \textit{ab initio} algorithm along $-\text{T}\leftrightarrow \Gamma \leftrightarrow \text{T}$, where the degeneracy is lifted along the entire path by breaking both $\mathcal{T}$ ({as well as} $\mathcal{PT}$) and $\sigma_v$.
The magnitude of the phonon energy splitting is presented in Fig.~\ref{fig:crystal_band_phonon}E, revealing a maximum splitting at the $\Gamma$ point and a local minimum splitting at T.

At $\Gamma$, the calculated phonon splitting is approximately 0.253 meV (2.05 cm$^{-1}$), which is in reasonable agreement with the experimental value of 1.27 cm$^{-1}$ ~\cite{CoSnS_PRL_2025}. 
Since the MBC can be strongly influenced by the position of the Fermi level, and a bias between DFT calculations and experimental measurements is common, arising from factors such as crystal imperfections (e.g., defects or unintentional doping), lattice expansion, and correlation-induced band renormalization--we further explored the dependence of the MBC on the Fermi level. To this end, we computed the splitting of the $E_g$ phonon mode at different Fermi-level positions. The detailed results are provided in the Supplementary Materials.

To further elucidate the lattice dynamics of $E_g$ induced by MBC, we calculate both the pseudo-angular momentum ($l_{ph}$) and the $z$-component of the angular momentum ($l_{z,\nu \bm{q}}$) along the $C_3$-invariant path, as shown in Fig.~\ref{fig:crystal_band_phonon}F. We note the angular momentum only has $z$ component due to the restriction of $C_3$ symmetry~\cite{Zhang2025_CP}. 
Both quantities exhibit a quantized value of unity, with sign reversals resulting from band inversion along the path, which fundamentally governs the emergence of the Weyl phonon~\cite{zhang2018double,FeSi_hmiao,zhang2025new} and phonon Hall effect~\cite{Strohm2005_THE, Nagaosa2010_THE_QuantumMagnets,Tao2012_Berry_THE, Ideue2017_THE_multiferroics, Zhang2021_ATHE_VI3, Nagaosa2019_Berry_THE,
Niu2022_MolecularBerry, Xue2025_Extrinsic}.

The quantized unity of the $l_{\text{ph}}$ arises from the $C_3$ symmetry of the system, while the quantization of $l_{z,\nu \bm{q}}$ can be attributed to two factors: (i) {At $\Gamma$,} $E_g$ is contributed solely by the vibration of S atoms residing at $C_3$-invariant Wyckoff positions~\cite{zhang2022chiral,Zhang2025_CP}. As illustrated in Fig.~\ref{fig:crystal_band_phonon}G, these two $\mathcal{P}$-related S atoms rotate in the same direction but with a $\pi$ phase difference at the $\Gamma$. (ii) {Away from $\Gamma$, } their energies are well-separated from other modes, resulting in minimal hybridization with other phonon modes.
It should be noted that the phonon splitting induced by MBC is not uniquely determined by $l_{z,\nu\bm{q}}$. As shown in Figs.~\ref{fig:crystal_band_phonon}E and ~\ref{fig:crystal_band_phonon}F, $l_{z,\nu\bm{q}}$ remains close to unity along the $C_3$-invariant path, yet the splitting of the $E_g$ modes varies significantly with a maximum at $\Gamma$.

\subsubsection{MBC-modified $E_u$ mode}

Figure~\ref{fig:crystal_band_phonon}H displays the MBC-modified phonon spectra of the $E_u$ mode along the path $-\text{T} \leftrightarrow \Gamma \leftrightarrow \text{T}$, where the MBC also lifts the degeneracy (dashed line) across the entire path. The magnitude of the phonon energy splitting for the $E_u$
modes is presented in Fig.~\ref{fig:crystal_band_phonon}I, revealing a local maximum splitting in the vicinity of T and local minimum splitting in the vicinity of $\Gamma$.
Both $E_{g/u}$ modes exhibit a broad distribution of splitting across the entire BZ, rather than being concentrated solely at the high-symmetry momentum. This widely distributed phonon symmetry breaking may give rise to observable transport signatures associated with phonon chirality.

At the $\Gamma$ point, the splitting calculated from our algorithm is about 0.002 meV (0.016 cm$^{-1}$), which is smaller than the experimentally observed value of 0.3 cm$^{-1}$ reported in the optical conductivity measurements~\cite{CoSnS_CP_2025}. 
This discrepancy suggests that the splitting observed in experiments is not primarily caused by the MBC. Combined with the asymmetric line shapes in experiments, we attribute the larger experimental splitting to the Fano resonance~\cite{Fano_1961}, which may arise from quantum interference between electronic and phonon excitations when Weyl cones approach the Fermi level~\cite{cappelluti_charged2012,CoSnS_CP_2025}. The absence of an asymmetric line shape in the $E_g$ mode further supports this interpretation, indicating that the phonon splitting mechanisms for the $E_g$ and $E_u$ modes are distinct.

We note that when the phonon peak exhibits an obvious asymmetric line shape, the Fano resonance correction must be taken into account, where the spectral peak position occurs at $\omega=\omega_0 + \frac{\Gamma}{2q}$. $\omega_0$ and $\Gamma$ denote the phonon frequency and the phonon linewidth, and $q$ is the Fano factor.
For Co$_3$Sn$_2$S$_2$, Ref.~\cite{CoSnS_CP_2025} reports that one of the $E_u$ modes has $1/q \approx 0$, indicating the absence of a noticeable Fano resonance and a peak position at $\omega=\omega_0$. In contrast, the other mode has $1/q = 0.775$, which leads to a peak shift of $\frac{\Gamma}{2q}$. Our calculations yield a corresponding phonon linewidth of $\Gamma = 0.095~\mathrm{meV}$. Using these parameters, the Fano-resonance-induced splitting of the $E_u$ mode is estimated to be $0.298~\mathrm{cm}^{-1}$, in good agreement with the experimentally measured value of $0.3~\mathrm{cm}^{-1}$ after further considering the MBC-induced value of $0.002~\mathrm{cm}^{-1}$. Further details are provided in the Supplementary Materials.

Figure~\ref{fig:crystal_band_phonon}J displays the distributions of $l_{ph}$ and $l_{z,\nu \bm{q}}$ for the $E_u$ modes, where $l_{z,\nu \bm{q}}$ is not quantized to unity. 
This arises because the $E_u$ mode involves both Co and S vibrations, and the Co atoms are not located at $C_3$-invariant Wyckoff positions~\cite{Zhang2025_CP}, as illustrated in Fig.~\ref{fig:crystal_band_phonon}K.

\subsection{Phonon splitting and effective magnetism from MBC}

\begin{figure}[h]
    \centering
    \includegraphics[width=0.47\textwidth]{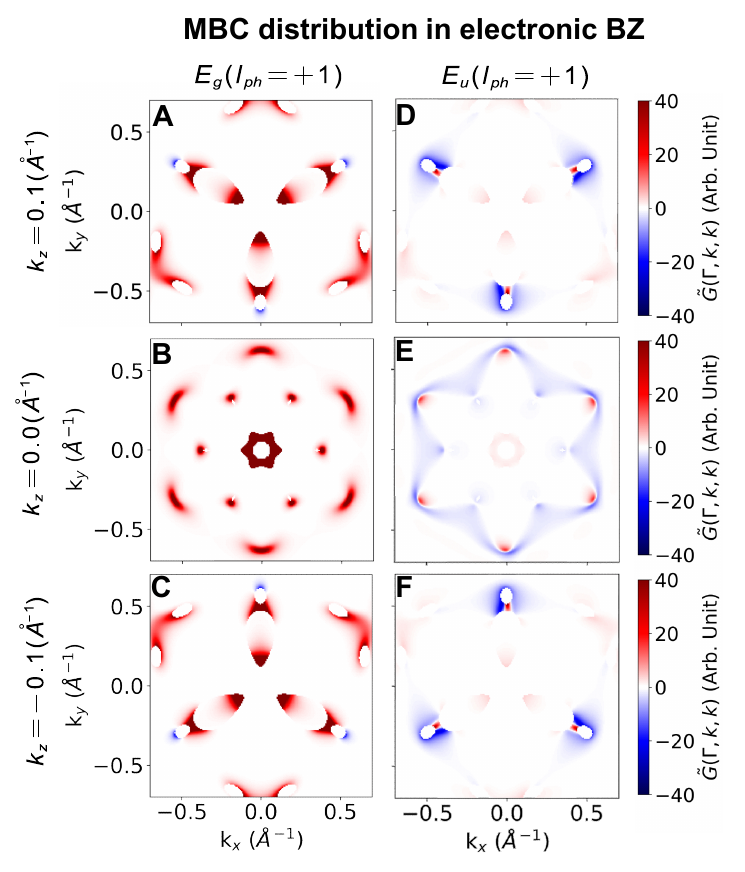}
    \caption{
    \textbf{Distribution of the MBC in the electronic BZ}. (A-C) are MBC distributions for the $E_g$ phonon mode with $l_{ph} = +1$ on three distinct $k_z$ planes.
    The MBC distribution for the $E_g$ mode with $l_{ph} = -1$ is of opposite sign. (D-F) are MBC distributions for the $E_u$ mode {with $l_{ph} = +1$. The $E_u$ mode with $l_{ph} = -1$ also exhibits a sign reversal. The MBC magnitude is significantly smaller for the $E_u$ modes, which consequently results in a smaller calculated phonon splitting.}}
    \label{fig:Gmatrix}
\end{figure}

For \CSS{}, the energy splitting of the $E_u$ mode is significantly smaller than that of the $E_g$ mode {at $\Gamma$ point}, a result captured by both our \textit{ab initio} calculations and experimental measurements~\cite{CoSnS_PRL_2025, CoSnS_CP_2025}. 
The MBC at $\Gamma$ can be expressed through the electron-phonon coupling constants (see Supplementary Materials for details), as
\begin{equation}
    \begin{aligned}
    \Tilde{G}_{\kappa \alpha, \kappa'\beta}(\bm{\Gamma}) &=  \frac{i}{N_k} \sum_{k v c}
        \frac{ [M^{\kappa \alpha \bm{\Gamma}}_{c\bm{k}, v\bm{k}}]^* M^{\kappa' \beta \bm{\Gamma}}_{ c \bm{k}, v \bm{k}}} {(\varepsilon_{c\bm{k} }-\varepsilon_{v\bm{k}})^2}-  
        \frac{i}{N_k} \sum_{k v c}
        \frac{ [M^{\kappa \alpha \bm{\Gamma}}_{v\bm{k}, c\bm{k}, }]^* M^{\kappa' \beta \bm{\Gamma}}_{v \bm{k}, c \bm{k}} } {(\varepsilon_{v\bm{k}}-\varepsilon_{c\bm{k}})^2} \\
        & = \sum_{k} \Tilde{G}(\bm{\Gamma}, \bm{k}, \bm{k}),
 \end{aligned}
\end{equation}
, i.e., $\tilde{G}(\bm{\Gamma}, \bm{k}, \bm{k})$ denotes the contribution of each electron–hole pair to the MBC. To explain their distinct magnitude of the energy splitting, we calculate the $\tilde{G}(\bm{\Gamma}, \bm{k}, \bm{k})$ for the $E_g$ and $E_u$ modes, the distribution across the whole electron BZ is shown in Fig.~\ref{fig:Gmatrix}.
The results indicate that $\tilde{G}_{E_u}(\bm{\Gamma}, \bm{k}, \bm{k})$ is significantly smaller in magnitude than $\tilde{G}_{E_g}(\bm{\Gamma}, \bm{k}, \bm{k})$, consistent with our earlier findings on MBC-induced phonon splitting. 
Furthermore, we observe that the distribution of $\tilde{G}_{E_u}(\bm{\Gamma}, \bm{k}, \bm{k})$ exhibits substantial cancellation between positive and negative contributions, which further explains the markedly smaller MBC-induced splitting of the $E_u$ mode.

We note that although the $E_u$ modes involves vibrations from both Co and S atoms, the atom-resolved MBC shows that S atom vibrations dominate its contribution (details are in the Supplementary Material Fig.~\ref{fig:Eu_seperate}). 
Combined with Figs.~\ref{fig:crystal_band_phonon}E and ~\ref{fig:crystal_band_phonon}I, it can be observed that while the parity of the $E_{g/u}$ modes remains unchanged at both $\Gamma$ and T, the relative phase between the two S atoms shifts by $\pi$. For instance, the $E_g$ mode exhibits a relative phase of $\pi$ at $\Gamma$ but 0 at T, whereas the $E_u$ mode shows a relative phase of 0 at $\Gamma$ and $\pi$ at T.
The distribution of the maximum phonon splitting for the $E_{g/u}$ modes switches along the $\Gamma$-T path, a reversal that correlates with the relative phase between two S atoms of the modes.

As the phonons undergo splitting due to time-reversal symmetry breaking, one may generally anticipate an effective nonzero phonon magnetic moment $M_{\rm ph}$, leading to a Zeeman-like relation $\Delta \omega = M_{\rm ph} \cdot H$. This suggests that the phonon magnetic moment can, in principle, be probed via an external magnetic field.
{Because the splitting of the $E_g$ mode is significantly larger than that of the $E_u$ mode, one may be tempted to infer that the $E_g$ mode has a larger phonon magnetic moment $M_{\mathrm{ph}}$.}  However, the direct coupling between ionic motion and the magnetic field is typically weak, so large phonon frequency splittings mainly arise from $\mathcal{T}$ breaking in the electronic system, either spontaneous or field-induced. In Co$_3$Sn$_2$S$_2$, the electronic magnetization is already saturated at zero field; as a result, the $E_g$ phonon splitting does not increase with applied magnetic field, consistent with the experimental observations in Ref.~\cite{CoSnS_PRL_2025}. {Thus, we believe that the relationship between phonon splitting and the phonon magnetic moment remains an open question.}

\section{Discussion}

In summary, we establish an MBC-based \textit{ab initio} framework that captures the electronic-order-induced symmetry breaking {in the lattice dynamic} and effective phonon magnetism in magnetic materials, as exemplified by Co$_3$Sn$_2$S$_2$. We find that the energy scale of circularly polarized phonons can coincide with that of chiral Weyl fermions, opening a pathway to chiral boson-fermion coupling and enabling control of topological states via phononic and magnetic interactions~\cite{hernandez2023observation}.
Our framework provides an efficient and accurate approach for modeling magnetic lattice dynamics, is readily extendable to full Brillouin-zone calculations, and facilitates the exploration of emergent collective phenomena. 

Beyond Co$_3$Sn$_2$S$_2$, our algorithm also predicts ferromagnetic materials such as FeCo and Co$_2$MnSi, both with $O_h$ symmetry, to exhibit FM-order-induced phonon splitting and effective phonon magnetism in their lattice dynamics. The MBC-induced maximum splitting along the $C_4$-invariant path is calculated to be 1.86 cm$^{-1}$ for FeCo and 0.16 cm$^{-1}$ for Co$_2$MnSi. 
Further details are provided in the Supplementary Materials. We note that larger splittings may be observed experimentally due to additional effects such as Fano resonances or electronic correlations.
Based on our results, we further identify design principles for achieving large effective phonon magnetic moments and sizable phonon splittings, which require strong electron-phonon coupling, pronounced spin-orbit coupling, and phonon modes with finite angular momentum.

\section{Methods}

The electronic structure and the density functional perturbation theory (DFPT) calculation is performed by the Quantum Espresso package~\cite{QE1, QE2, QE3}. 
The norm-conserving pseudo-potential of the Perdew-Burke-Ernzerhof (PBE) exchange-correlation functional~\cite{perdew_generalized_1996, dojo, ONCV_Hamann} was used. 
The kinetic energy cutoff is set to 90 Ry. The Brillouin zone was sampled using a $8\times8\times8$ Monkhorst-Pack mesh. {The force constants are obtained by Fourier transforming the dynamical matrices calculated by DFPT on a $3\times 3\times3$ $q$-mesh.}
The MBC is calculated under the first-order-perturbation framework and calculated by the Wannier-based electron-phonon coupling  vortex with modified EPW code~\cite{mostofi2008wannier90,mostofi2014updated,EPW1,EPW2,EPW3}.
The MBC is Wannier-interpolated on a $80\times80\times80$ fine $k$-mesh.
According to the Fermi level extracted from STM experiments~\cite{Yin2019_CSS_Negative}, we shift the Fermi level by 0.025 eV to align it with the flat band along the $\Gamma$--$K$ direction.
The core code will be published along with the manuscript.


\clearpage 

%
\bibliography{main} 
\bibliographystyle{sciencemag}

%
%
%
%
%
%


\section*{Acknowledgments}

We acknowledge the helpful discussion with Haijun Zhang, Fuyi Wang, Xi Dai, Yang Gao, Yafei Ren, Shang Ren, Run Yang and Martin Dressel. 

\paragraph*{Foundings:} T. Z. acknowledge the support from National Key R\&D Project (grant Nos. 2023YFA1407400 and 2024YFA1409200), the National Natural Science Foundation of China Grant No. 12374165, and the Strategic Priority
Research Program (B) of the Chinese Academy of Sciences (CAS) (grant no. XDB1720000).

\paragraph*{Author contributions:}

T.Z. conceived of the presented idea. T.Z. and S.Z. developed the theory and performed the first-principles calculations. All authors discussed the results and contributed on the manuscript.

\paragraph*{Competing interests:} 

The authors declare that they have no competing interests.

\paragraph*{Data, Code, and Materials Availability:}

All data and code needed to evaluate and reproduce the results in the paper are present in the paper,  the Supplementary Materials, Zenodo repository (No. 19906198, \url{https://zenodo.org/records/19906198}) and/or the github repository \url{https://github.com/ZS137/MBC.}
This study did not generate new materials.

\end{document}